# Spin-Orbit Coupling Response Theory within the Oriented Spin Space Group Framework

Yuntian Liu[1], Xiaobing Chen[2,1], Yutong Yu[1], Jesús Etxebarria[3], J. Manuel Perez-Mato[3] and Qihang Liu[1,2,*]

[1]*State Key Laboratory of Quantum Functional Materials, Department of Physics, and Guangdong Basic Research Center of Excellence for Quantum Science, Southern University of Science and Technology (SUSTech), Shenzhen 518055, China*

[2]*Quantum Science Center of Guangdong–Hong Kong–Macao Greater Bay Area (Guangdong), Shenzhen 518045, China*

[3]*Faculty of Science and Technology, University of the Basque Country/Euskal Herriko, Unibertsitatea (EHU), Bilbao, Spain*

[*]Email: liuqh@sustech.edu.cn

**Abstract**

Spin-orbit coupling (SOC) plays a pivotal role in a broad range of phenomena in condensed matter physics. The recently developed spin space group theory offers a new perspective for characterizing the fundamental symmetry-breaking features of SOC-related phenomena, yet its full potential remains to be explored. Here, we provide a general framework that relates the SOC-induced symmetry breaking to modifications of physical properties and emergence of new physical effects. This approach incorporates the recently developed oriented spin space group (OSSG) formalism, which can explicitly describe the symmetry reduction induced by SOC. By parameterizing the SOC Hamiltonian using a SOC tensor that connects real and spin spaces, an expansion in terms of the SOC components can be constructed for the response tensor describing any physical quantity. This framework determines not only whether a SOC-induced response is symmetry allowed, but also the lowest order in SOC at which it emerges and its dependence on magnetic anisotropy. As representative applications, we analyze the magnetoelectric coupling in Type-II multiferroics and show that magnetically induced electric polarization in different systems can be the result of SOC effects of different orders, and can exhibit varying characteristics of magnetic anisotropy. We further apply this theory to the anomalous Hall effect and spin magnetization in antiferromagnets, demonstrating that their distinct transformation properties under the OSSG allow a large anomalous Hall response to coexist with weak spin magnetization. A high-throughput analysis of materials in the MAGNDATA database identifies several symmetry-based material classes favorable for this coexistence. Overall, the proposed framework provides a unified symmetry description of SOC-induced phenomena, and offers a systematic way for the identification and design of SOC-driven functional materials.

## I. INTRODUCTION

Spin-orbit coupling (SOC), a relativistic interaction that couples the spin and orbital degrees of freedom of electrons, plays a central role in modern condensed matter physics. In nonmagnetic systems, SOC lifts band-degeneracy, giving rise to spin splitting and topological properties [1-3]. In magnetic systems, magnetic order introduces an additional degree of freedom into the SOC interaction by favoring specific spin orientations relative to the crystal lattice, thereby giving rise to magnetic anisotropy, a key property in spintronics [4-6]. More broadly, the interplay between magnetic order and SOC underlies a variety of important phenomena, including the anomalous Hall effect (AHE) [7-10], spin-orbit torque [11-13], magnetoelectric coupling [14-16], and magnetic topological phases [17-20]. The richness of these phenomena, together with their diverse manifestations across different magnetic systems, has greatly broadened the applications of magnetic materials and devices.

From the theoretical perspective, symmetry provides a unifying framework for understanding the emergence, classification, and material realization of these SOC-driven responses. Traditionally, the symmetry breaking in magnetic systems has been formalized within the Landau theory of phase transitions [21]: starting from a high-symmetry paramagnetic phase, described by the crystallographic space group (SG) augmented by time-reversal symmetry ($T$), the formation of magnetic order drives the system into a low-symmetry phase described by a magnetic space group (MSG) [22-25], in which rotation operations act simultaneously on both lattice and spin [Fig. 1(a)]. In this scenario, SOC is implicitly present in any symmetry consideration, making it difficult to disentangle the symmetry-breaking contributions of magnetic order and SOC. Consequently, the specific role of SOC in many physical phenomena remains obscured. The AHE provides a representative example: although it is commonly associated with the time-reversal symmetry breaking induced by the magnetic order, SOC is essential for its emergence in a broad class of collinear and coplanar magnets [26,27]. To isolate the role of SOC, one must instead consider the symmetry of the magnetic system in the absence of SOC, as described by a spin space group (SSG) [26,28-34], which in general has a higher symmetry than the associated MSG. Thus, if this SSG is taken as a virtual intermediate symmetry group of the system, many SOC-driven phenomena of interest can be mathematically expressed in a perturbative approach with SOC as the symmetry-breaking perturbation.

A general theoretical description with SOC as a symmetry-breaking perturbation requires two ingredients: first, the transformation properties of the SOC interaction under the symmetry of the SOC-free magnetic system, as described by its SSG; second, the group-subgroup relations connecting this SOC-free SSG to the MSG that emerges once SOC is introduced [Fig. 1(b)]. The first question has been discussed in recent studies of the AHE in ferromagnetic (FM) systems [35,36], where the effect is microscopically driven by the anisotropy of multipole moments. This coupling represents a particular case because the FM order allows the SSG to be simplified to a spinless SG augmented by time-reversal symmetry [37]. For the more general case of any kind of magnetic structures, it is the recently developed oriented spin space group (OSSG) framework [38] that provides a suitable platform for characterizing the SOC-induced symmetry breaking. An OSSG retains all symmetry operations of the corresponding SSG while specifying the orientations of its spin-space operations relative to real space. It therefore describes the symmetry of a SOC-free system in the same manner as an SSG [39-45], while additionally using this orientational information to establish a group-subgroup correspondence with the associated MSG that emerges once SOC is introduced. This enables the establishment of a general framework to characterize SOC-induced effects across a broad range of physical phenomena and material systems.

In this work, we develop a symmetry-based formalism to compute response properties of SOC-induced phenomena using the OSSG framework. Specifically, by parameterizing the SOC Hamiltonian and imposing the symmetry constraint encoded by the OSSG, we derive the dependence of various physical quantities on both the SOC strength and the magnetic anisotropy (Sec. II). This framework enables the role of SOC in any physical response tensor to be determined and can be readily connected to established tensor theory through its description using generalized Jahn symbols [49]. To demonstrate the generality of our approach, we apply it to representative SOC-related phenomena in several systems, including magnetoelectric coupling in Type-II multiferroics and magnetization in spin-orbit magnets. Type-II multiferroics, whose electric polarization is induced by SOC, provide an ideal platform for applying the present approach. As representative examples, we show that in the ferromagnet (FM) $VSe_2$ and the spiral antiferromagnet (AFM) $VCl_2$, SOC-induced electric polarization can originate from different orders in SOC and exhibit distinct dependence on spin orientation (Sec. III). The predicted dependences of electric polarization on spin orientation are further verified by DFT calculations. In spin-orbit magnets, we perform a high-throughput analysis of 224 candidate materials in the MAGNDATA database and classify them into a few types of closely related magnetic structures based on their distinct SOC dependence on AHE and spin magnetization (Sec. IV).

## II. SYMMETRY BREAKING BY SPIN-ORBIT COUPLING

### a. Transformation and parameterization of the SOC Hamiltonian under OSSG symmetry

To describe SOC as a symmetry-breaking perturbation within the OSSG framework, we first examine the form of the SOC Hamiltonian. Within the MSG framework, since symmetry operations act simultaneously on both spin and spatial coordinates, the coordinate bases associated with the magnetic moments and the crystal lattice need not be distinguished [46]. Accordingly, the SOC Hamiltonian is naturally written in the conventional three-component dot-product form, $\hat{H}_{SOC} = \lambda(\hat{L}_1\hat{\sigma}_1 + \hat{L}_2\hat{\sigma}_2 + \hat{L}_3\hat{\sigma}_3)$. However, early studies of spin groups already emphasized the existence of an independent carrier space for spin degrees of freedom, namely, the spin space [28,29]. This implies that, in the absence of any prior locking between real space and spin space, these two spaces can be described by independent sets of unit basis vectors, $\boldsymbol{r}_i$ and $\boldsymbol{s}_j$, respectively. Therefore, to analyze SOC-induced symmetry breaking, the SOC Hamiltonian should first be expressed in a more general form defined over independent real and spin spaces:

$$\hat{H}_{SOC} = \lambda \hat{\boldsymbol{L}}^T \boldsymbol{\chi} \hat{\boldsymbol{\sigma}} = \lambda(\hat{L}_1 \quad \hat{L}_2 \quad \hat{L}_3)\begin{pmatrix} \boldsymbol{r}_1 \cdot \boldsymbol{s}_1 & \boldsymbol{r}_1 \cdot \boldsymbol{s}_2 & \boldsymbol{r}_1 \cdot \boldsymbol{s}_3 \\ \boldsymbol{r}_2 \cdot \boldsymbol{s}_1 & \boldsymbol{r}_2 \cdot \boldsymbol{s}_2 & \boldsymbol{r}_2 \cdot \boldsymbol{s}_3 \\ \boldsymbol{r}_3 \cdot \boldsymbol{s}_1 & \boldsymbol{r}_3 \cdot \boldsymbol{s}_2 & \boldsymbol{r}_3 \cdot \boldsymbol{s}_3 \end{pmatrix}\begin{pmatrix} \hat{\sigma}_1 \\ \hat{\sigma}_2 \\ \hat{\sigma}_3 \end{pmatrix}, \quad (1)$$

where $\lambda = \hbar/(mc)^2$, $\hat{\boldsymbol{L}} = \nabla V(\hat{\boldsymbol{q}}) \times \hat{\boldsymbol{p}}$ and $\hat{\boldsymbol{\sigma}}$ denote the SOC coefficient, the effective orbital angular momentum operator and the spin operator, respectively. Here, $\boldsymbol{\chi}$ is the tensor that connects the real-space basis vectors $\boldsymbol{r}_i$ and the spin-space basis vectors $\boldsymbol{s}_j$, forming a $3 \times 3$ matrix defined as $\boldsymbol{\chi} = \{\chi_{ij} = \boldsymbol{r}_i \cdot \boldsymbol{s}_j | i = 1,2,3; j = 1,2,3\}$. We call $\boldsymbol{\chi}$ the SOC tensor [35,47]. Eq. (1) is the starting point for analyzing the transformation properties of the SOC interaction under OSSG operations.

The OSSG framework explicitly specifies the spin orientation in real space and the corresponding MSG that emerges after SOC is introduced. The symmetry operations of an OSSG can be written in the general form $\{g_s||g_l|\tau\}$, where $g_s$ denotes a spin-space operation, including a spin rotation $U_{\boldsymbol{m}}(\varphi)$ and time-reversal symmetry $T$; $g_l$ denotes a real-space operation, including a lattice rotation $C_{\boldsymbol{n}}(\theta)$ and space-inversion symmetry $P$; and $\tau$ represents a translation. The rotation axes for spin and lattice rotations, $\boldsymbol{m}$ and $\boldsymbol{n}$, are expressed in the real-space basis, and $\varphi$ and $\theta$ are the corresponding rotation angles. When SOC is introduced, the locking between spin-space and real-space rotations results in that all OSSG operations with $\boldsymbol{m} \neq \boldsymbol{n}$ or $\varphi \neq \theta$ are broken, reducing the group to its MSG subgroup that allows the SOC-related effects.

For a general OSSG operation $\{g_s||g_l|\tau\}$, the transformation of the SOC Hamiltonian can be expressed as:

$$\hat{\rho}^{-1}_{\{g_s||g_l|\tau\}} \lambda \hat{\boldsymbol{L}}^T \boldsymbol{\chi} \hat{\boldsymbol{\sigma}} \, \hat{\rho}_{\{g_s||g_l|\tau\}}$$
$$= \lambda det(R_s) det(R_l) [\hat{\boldsymbol{L}}^T R_l^{-1}] \boldsymbol{\chi} [R_s \hat{\boldsymbol{\sigma}}]$$
$$= \lambda \hat{\boldsymbol{L}}^T det(R_s) det(R_l) [R_l^{-1} \boldsymbol{\chi} R_s] \hat{\boldsymbol{\sigma}}, \quad (2)$$

where $\hat{\rho}_{\{g_s||g_l|\tau\}}$ is the representation operator of $\{g_s||g_l|\tau\}$ in Hilbert space; $R_l$ and $R_s$ are the 3×3 representation matrices in the three dimensional (3D) Euclidean space corresponding to the rotation parts of $g_l$ and $g_s$, respectively. The factors $det(R_l)$ and $det(R_s)$, each taking the value -1 or 1, depend on whether $R_l$ contains the space-inversion operation and whether $R_s$ contains the time-reversal operation, respectively. Therefore, the transformation of the SOC term under an OSSG operation can be characterized by the SOC tensor $\boldsymbol{\chi}$, based on the following transformation rule:

$$\boldsymbol{\chi} \xrightarrow{\{g_s||g_l|\tau\}} \boldsymbol{\chi}' = det(R_s)det(R_l)R_l^{-1}\boldsymbol{\chi}R_s. \quad (3)$$

Because $\boldsymbol{\chi}$ transforms under OSSG operations in the same manner as the SOC term, the symmetry breaking induced by $\boldsymbol{\chi}$ is mathematically equivalent to that driven by the full SOC term in the Hamiltonian. Therefore, $\widehat{H}_{SOC}$ can be replaced with $\lambda\boldsymbol{\chi}$ in the following analysis, where $\lambda$ conveys the strength of the SOC. This can be viewed as a reduction from a Hilbert-space representation to one defined in a 3D Euclidean space, thereby substantially simplifying the subsequent symmetry analysis.

**b. Expansion of property tensors in powers of the SOC strength**

Building on the above simplification, we now discuss the relation between SOC-induced effects and $\lambda\boldsymbol{\chi}$ by analyzing the symmetry constraints imposed on this tensor by the OSSG. According to the representation compatibility principle, a response is symmetry-allowed when the symmetry-breaking perturbation and the resulting physical quantity transform according to compatible representations. This principle is widely used to analyze symmetry breakings induced by strain or external fields; here, it is extended to the intrinsic SOC and the associated physical quantities. Specifically, when the OSSG is not equivalent to an MSG, SOC breaks certain symmetries and thus transforms according to a nonidentity representation $\mathcal{R}_\chi$ of the OSSG. Similarly, a SOC-induced physical quantity represented by a tensor of rank $r$ $A_{a1,\ldots ar}$ transforms according to a nonidentity representation $\mathcal{R}_A$ of the OSSG. If the $i$-fold direct product representation $\mathcal{R}_\chi^{\otimes i}$ contains $\mathcal{R}_A$, then $A_{a1,\ldots ar}$ can emerge at $i$th order in $\lambda\boldsymbol{\chi}$. In general, the relation between $A_{a1,\ldots ar}$ and $\lambda\boldsymbol{\chi}$ can be expressed as a series expansion:

$$A_{a1,\ldots ar}[\boldsymbol{\chi}] = \sum_{ij}\omega^{A,(1)}_{a1,\ldots ar,ij}\lambda\chi_{ij} + \sum_{ijkl}\omega^{A,(2)}_{a1,\ldots ar,ij,kl}\lambda^2\chi_{ij}\chi_{kl} + \cdots, \quad (4)$$

where $\boldsymbol{\omega}^{A,(n)}$ is a rank-$(2n+r)$ response tensor that relates the $n$th-order products of $\lambda\boldsymbol{\chi}$ to $A_{a1,\ldots ar}$, and $i$, $j$, $k$, $l$,... label the tensor components. The allowed components of $\boldsymbol{\omega}^{A,(n)}$ are dictated by the OSSG symmetry.

Therefore, once the OSSG-allowed form of $\boldsymbol{\omega}^{A,(n)}$ is derived at each order for a given material, the dependence of $A_{a1,\ldots ar}$ on $\lambda\boldsymbol{\chi}$ can be established. This procedure can also be formulated in terms of the Jahn symbols [48,49], which characterize the symmetry properties of physical tensors and have recently been extended to SSGs [49]. Following the generalized Jahn symbol notation for tensor properties, $V$ denotes a polar-vector index in real space, $M$ denotes a polar-vector index with respect to spin-space operations, $e$ denotes odd parity factor under space inversion, and $a$ denotes odd parity factor under time reversal. Accordingly, $\boldsymbol{\chi}$ is of type $aeVM$. Square or curly brackets are used to denote symmetry or antisymmetry between pairs of indices. If the physical quantity $A_{a1,\ldots ar}$ is of type $J^A$, the corresponding response tensor $\boldsymbol{\omega}^{A,(n)}$ has the Jahn symbol $J^A \cdot (aeVM)^n$. This formulation readily connects SOC response theory with established tensor theory, thereby demonstrating its broad applicability to describe various physical responses.

After the response tensor $\boldsymbol{\omega}^{A,(n)}$ has been determined, we only need to substitute the specific $\boldsymbol{\chi}$ corresponding to the actual magnetic ground state of the material under SOC to obtain the expression for $A_{a1,\ldots ar}$ in that ground state. This is precisely where the convenience of the OSSG framework becomes apparent: following OSSG standardization, the axes of the spin-space and real-space operations are expressed on a common coordinate basis, i.e., $\boldsymbol{r}_i = \boldsymbol{s}_i$. As a result, the SOC tensor $\boldsymbol{\chi}$ in the ground state retains only the diagonal components, $\chi_{ij} = \delta_{ij}$, as shown in Fig. 2(a). On the other hand, different choices of $\boldsymbol{\chi}$ can be used to elucidate the effects of magnetic anisotropy [47]. When the spins undergo collective rotation, this rotation is applied to the spin-space coordinate basis $\boldsymbol{s}_j$ relative to the real-space coordinate basis $\boldsymbol{r}_i$, rather than to the spins themselves with respect to $\boldsymbol{s}_j$, as shown in Fig. 2(b). The OSSG restrictions on the response tensors $\boldsymbol{\omega}^{(n)}$ between the physical quantity $A_{a1,\ldots ar}$ and the SOC tensor $\boldsymbol{\chi}$ are in fact valid for any orientation of the spin basis with respect to the lattice basis, with $\boldsymbol{\chi}$ describing in this case the relation between the two bases, rather than simply being the identity matrix. In this way, by choosing different orthogonal transformation matrices for $\chi$, one can determine the influence of magnetic anisotropy on any physical quantity.

**c. Relation to other symmetry analyses based on magnetic group symmetry**

The present framework complements conventional symmetry analyses based on MSGs, by adopting a different symmetry-breaking variable. In conventional approaches, such as the Landau theory of magnetism [46], the magnetic configuration $\boldsymbol{M}(\boldsymbol{r}_i)$, with the spins locked to the lattice, serves as the symmetry-breaking variable, and the parent symmetry is taken to be the nonmagnetic SG augmented by time-reversal symmetry [Fig. 1(a)]. In the present approach, the symmetry breaking is considered in two steps, with a first step being the symmetry breaking into the SOC-free SSG, with a fixed orientation with respect to the lattice as described by the OSSG, and in a second step, the effect of the SOC is introduced as an additional symmetry-breaking perturbation. In this way, the formalism can directly identify the symmetry-allowed dependence of physical properties on SOC, including the lowest order at which a given SOC-induced effect can appear.

In addition, this framework eliminates the need to choose the local magnetic-moment order parameters on a case-by-case basis according to the specific Wyckoff positions of magnetic atoms in different materials. This feature enables systematic analyses of SOC effects using the existing magnetic materials databases. Furthermore, in contrast to the previous consensus that spin-group theory can describe only SOC-free effects, this framework shows that spin groups are also capable of describing SOC-related effects, thereby offering insights beyond those provided by MSGs. In the following, we demonstrate the application of this approach through two examples: magnetoelectric coupling in type-II multiferroics and comparison between spin magnetization and AHE in AFMs.

## III. MAGNETOELECTRIC COUPLING IN TYPE-II MULTIFERROIC

**SYSTEMS**

Magnetoelectric multiferroic systems are typically classified as Type-I or Type-II, depending on whether magnetic ordering and electric polarization have independent or related origins [50]. Type-II multiferroics are magnetic insulators in which electric polarization is induced by magnetism, possessing a nonpolar parent SG in the paramagnetic phase but a polar MSG in the magnetic phase. Within this classification, it is meaningful to further distinguish the specific role of SOC, for which the SSGs combined with SOC response theory provides a natural framework. Thus, a polar SSG indicates that the nonpolar-to-polar symmetry breaking originates from the magnetic order rather than SOC. In such systems, the polarization typically originates from exchange striction, with representative examples including $HoMnO_3$ and $GdMn_2O_5$ [51-53]. By contrast, a nonpolar SSG indicates that the emergence of electric polarization requires SOC. SOC response theory can further identify the lowest order in SOC at which the polarization appears and determine how it varies with the spin orientation.

**a. Formulation of electric polarization**

Macroscopic electric polarization $\boldsymbol{P}$ is a canonical polar vector in real space transforming according to only point group operations. Specifically, it transforms as a vector under real-space rotation operations, reverses sign under space-inversion, and remains invariant under time-reversal and spin-space rotation operations. Therefore, the transformation of $\boldsymbol{P}$ under a spin point group operation $\{g_s||g_l\}$ can be expressed as:

$$\boldsymbol{P} \xrightarrow{\{g_s||g_l\}} R_l\boldsymbol{P}. \tag{5}$$

In terms of the Jahn symbol description, $\boldsymbol{P}$ transforms as $V$. The relationship between $\boldsymbol{P}$ and the SOC tensor $\lambda\boldsymbol{\chi}$, is then expressed, following Eq. (4), as a series expansion:

$$P_a[\boldsymbol{\chi}] = \omega_a^{P,(0)} + \lambda \sum_{ij} \omega_{a,ij}^{P,(1)} \chi_{ij} + \lambda^2 \sum_{ijkl} \omega_{a,ij,kl}^{P,(2)} \chi_{ij}\chi_{kl} + \cdots, \tag{6}$$

where $\boldsymbol{\omega}^{P,(n)}$ denotes an undetermined tensor of rank $(2n+1)$; $a$ indicates the components of the vector $\boldsymbol{P}$; the indices $i$, $j$, $k$, $l$ label the components of the SOC tensor. Here, the zeroth-order term $\omega_a^{P,(0)}$ is independent of $\lambda\boldsymbol{\chi}$ and is directly constrained by the SSG symmetry, representing the value of $P_a$ in the absence of SOC. Thus, a nonzero $\omega_a^{P,(0)}$ corresponds to the subclass of Type-II multiferroics with a polar SSG, in which at least part of the induced electric polarization does not require SOC, as exemplified by $HoMnO_3$ and $GdMn_2O_5$ [51-53]. The invariance condition for the transformation by the point-group part $\{g_s||g_l\}$ of an operation of the OSSG requires the relation between $\boldsymbol{P}$ and $\lambda\boldsymbol{\chi}$ to satisfy:

$$R_l\,\boldsymbol{P}[\boldsymbol{\chi}] = \boldsymbol{P}[det(R_s)det(R_l)R_l\boldsymbol{\chi}R_s^{-1}]. \quad (7)$$

By combining Eqs. (6) and (7), the allowed form of the tensor $\boldsymbol{\omega}^{P,(n)}$ can be determined for each order of the expansion.

We illustrate the specific symmetry of $\boldsymbol{\omega}^{P,(1)}$ as an example. The constraint can be written as:

$$\omega_{nmi}^{P,(1)\prime} = \sum_{kjl} det(R_s)det(R_l)[R_l]_{nk}[R_l]_{mj}[R_s]_{il}\omega_{kjl}^{P,(1)}, \quad (8)$$

which can be easily derived using Jahn symbols. Starting from Eqs. (3) and (5), since $\boldsymbol{P}$ is of type $V$ and $\boldsymbol{\chi}$ is of type $aeVM$, it follows that $\boldsymbol{\omega}^{P,(1)}$ must be described by the product $V\cdot aeVM$, i.e., $aeV^2M$, which is precisely Eq. (8). Similarly, the constraint on the coefficient tensor $\boldsymbol{\omega}^{P,(2)}$ of the second-order term of $\boldsymbol{\chi}$ is given by:

$$\omega_{nmipq}^{P,(2)\prime} = \sum_{kjlst}[R_l]_{nk}[R_l]_{mj}[R_s]_{il}[R_l]_{ps}[R_s]_{qt}\omega_{kjlst}^{P,(2)}. \quad (9)$$

Again, the properties of transformation for $\boldsymbol{\omega}^{P,(2)}$ are fully encoded in the generalized Jahn symbol $V\cdot aeVM\cdot aeVM$, that is, $V^2MVM$, which is equivalent to Eq. (9). Generally, the transformation for $\boldsymbol{\omega}^{P,(n)}$ corresponds to the Jahn symbol $V\cdot(aeVM)^n$. In the following, we consider two materials, i.e., the FM $VSe_2$ and the spiral AFM $VCl_2$, as examples to demonstrate how the electric polarization of realistic materials can be analyzed by applying the above general expansion.

**b. Type-II multiferroic materials examples: $VSe_2$ and $VCl_2$**

$VSe_2$ is a quasi-two-dimensional in-plane collinear FM [54,55]. The crystal structure and magnetic configuration are shown in Fig. 3(a). Its ferroelectric polarization is believed to be driven by a SOC-induced spin-dependent variation in the *p*-*d* hybridization [56]. The symmetry of the magnetic structure of $VSe_2$ is given by the OSSG $P^1\text{-}6^1m^12^{\infty_{100}m}1$ (here we adopt the Chen-Liu symbol from Ref. [26]). Its MSG is $Am'm2'$, with the orthorhombic unit-cell parameters being **c**, **a** and **a**+2**b** with respect to the hexagonal unit cell. A set of generators of the corresponding oriented spin point group are $\{1||\text{-}6^1_{001}\}$, $\{1||m_{100}\}$, $\{m_{001}||1\}$ and $\{\infty_{100}||1\}$. Here we choose the unit basis vectors $\boldsymbol{r}_i$ and $\boldsymbol{s}_j$ $(i,j=1,2,3)$ of real space and spin space to share the same conventional orthogonal crystal orientation $\boldsymbol{r}_1 \parallel \boldsymbol{s}_1 \parallel [100]$, $\boldsymbol{r}_2 \parallel \boldsymbol{s}_2 \parallel [120]$ and $\boldsymbol{r}_3 \parallel \boldsymbol{s}_3 \parallel [001]$.

Next, we derive the response of $\boldsymbol{P}$ to the SOC tensor $\lambda\boldsymbol{\chi}$ based on the OSSG constraints resulting from Eq. (7). We first consider the coefficient $\boldsymbol{\omega}^{P,(0)}$ of the zeroth-order SOC term, with Jahn symbol *V*. Its three components change under the operation $\{1||\text{-}6^1_{001}\}$, whereas it should remain invariant for any operation of the spin point group. Consequently $\boldsymbol{\omega}^{P,(0)}$ is forced to vanish, as expected from the nonpolar character of the OSSG, and $\boldsymbol{P}$ is induced entirely by the SOC. We then consider the coefficient $\boldsymbol{\omega}^{P,(1)}$ of the first-order SOC term, which is of type $aeV^2M$. $\boldsymbol{P}$ remains invariant under the spin-only operations $\{m_{001}||1\}$ and $\{\infty_{100}||1\}$, whereas none of the components of the linear term of $\boldsymbol{\chi}$ can remain invariant under these operations. Therefore, all components of $\boldsymbol{\omega}^{P,(1)}$ must also vanish. In general, due to the spin-only group $^{\infty m}1$, any SOC-induced electric polarization in collinear magnets must originate from at least second-order SOC terms, independently of the lattice type.

For the second-order SOC contribution, we impose the OSSG invariance conditions on the response tensor $\boldsymbol{\omega}^{P,(2)}$ (type $V^2MVM$) and substitute the SOC tensor $\boldsymbol{\chi}$ corresponding to the magnetic ground state. In the standardized OSSG setting, $\boldsymbol{\chi}$ in the ground state retains only its diagonal components, $\chi_{ij}=\delta_{ij}$. The resulting second-order polarization is therefore given by [57]:

$$P_1^{(2)}=0, \qquad P_2^{(2)}=\lambda^2[\omega^{P,(2)}_{1,11,21}-\omega^{P,(2)}_{1,12,22}], \qquad P_3^{(2)}=0. \tag{10}$$

Thus, $\boldsymbol{P}$ is oriented along the $[120]$ direction, consistent with the symmetry constraints imposed by the corresponding MSG. Moreover, its explicit $\lambda^2$ dependence demonstrates that polarization arises as a second-order effect in SOC.

This SOC response formalism also enables a direct description of how a collective rotation of the spins modifies $\boldsymbol{P}$. As discussed in Sec. II, this can be achieved simply by choosing $\chi$ as the corresponding three-dimensional orthogonal transformation matrix. Below, we consider collective spin rotations about the [001] and [120] axes, which encompass several common spin configurations aligned with high-symmetry crystallographic directions. For a collective rotation of the spins about the [001] axis, the corresponding variation of $\boldsymbol{P}$ can be expressed as [57]:

$$P_1^{(2)} = \lambda^2[\omega_{1,11,21}^{P,(2)} - \omega_{1,12,22}^{P,(2)}]sin(2\varphi), \quad P_2^{(2)} = \lambda^2\left[\omega_{1,11,21}^{P,(2)} - \omega_{1,12,22}^{P,(2)}\right]\cos(2\varphi), \tag{11}$$

where $\varphi$ is the angle between spin orientation and the [100] direction. Therefore, when the spins rotate about the [001] axis, the magnitude of $\boldsymbol{P}$ arising from the lowest-order SOC contribution remains unchanged, while the vector direction rotates within the plane with half the period (according to $2\varphi$), as shown in Fig. 3(b). By contrast, when the spins collectively rotate about the [120] axis, the corresponding variation of $\boldsymbol{P}$ can be expressed as [57]:

$$P_1^{(2)} = 0, \qquad P_2^{(2)} = \lambda^2\left[\omega_{1,11,21}^{P,(2)} - \omega_{1,12,22}^{P,(2)}\right]\cos^2(\phi), \tag{12}$$

where $\phi$ is the angle between spin orientation and the [100] direction. Thus, during the collective spin rotation about the [120] axis, the lowest-order SOC contribution to $P_1^{(2)}$ along the [100] direction remains zero, whereas the dominant component $P_2^{(2)}$ lies along the [120] direction and varies as $\cos^2(\phi)$, as shown in Fig. 3(c).

The above analysis of the variation of electric polarization under collective spin rotation in $VSe_2$, based solely on the symmetry information from OSSG, captures the essential features of its magnetoelectric coupling and provides guidance for experimentally controlling the polarization direction through spin orientation. This approach reveals information that cannot be obtained from the MSG alone. For example, when the spins are oriented at an angle of π/4 relative to the [100] direction, the $P_2$ component vanishes even though the low-symmetry magnetic point group $m'$ (perpendicular to [001]) for this spin orientation allows a non-zero $P_2$. Thus, for this orientation only SOC effects of higher order can induce a non-zero component along the [120] direction.

For the non-collinear spiral AFM $VCl_2$ system, the situation is quite different. $VCl_2$ has a quasi-two-dimensional layered structure with intralayer in-plane spin spiral configuration and interlayer antiparallel spin arrangement [58,59], as shown in Fig. 3(d). Correspondingly, the magnetic symmetry of $VCl_2$ is given by the OSSG $P^{2_{001}}\text{-}3^{2_{001}}m^{1}1|(3^{1}_{120},3^{1}_{120},2_{120})^{m_{120}}1$ and the MSG $C_cc$ with the monoclinic axis along [100] in the trigonal setting. Operations $\{3^{1}_{120}||1\}$ and $\{2_{120}||1\}$, are derived from the point group parts of the spin translation group, which ensure the intralayer spiral and interlayer antiparallel spin geometries, respectively. The unit basis vectors $\boldsymbol{r}_i$ and $\boldsymbol{s}_j$ $(i,j=1,2,3)$ of real space and spin space are also chosen as the same conventional orthogonal crystal orientation $\boldsymbol{r}_1 \parallel \boldsymbol{s}_1 \parallel [100]$, $\boldsymbol{r}_2 \parallel \boldsymbol{s}_2 \parallel [120]$ and $\boldsymbol{r}_3 \parallel \boldsymbol{s}_3 \parallel [001]$. Different from the case of $VSe_2$, since the constraints imposed by a collinear magnetic structure no longer apply, a linear SOC effect on the electric polarization is allowed in $VCl_2$. $\boldsymbol{P}$ of the ground state can be shown to be given by [57]:

$$P_1^{(1)}=0,\qquad P_2^{(1)}=\lambda\omega_{1,12}^{P,(1)},\qquad P_3^{(1)}=0. \tag{13}$$

Accordingly, $\boldsymbol{P}$ in $VCl_2$ depends linearly on SOC rather than arising only at higher orders. As a result, it can remain appreciable even in systems with relatively weak SOC. Interestingly, its orientation lies along the [120] direction, whereas the MSG constraints allow polarization throughout the entire (100) plane. As in the previous case, this shows that different polarization components can originate from SOC contributions of different orders. In fact, following the same procedure and retaining terms up to $\boldsymbol{\omega}^{P,(3)}$, it can be shown that a non-zero component $P_3^{(3)}\neq 0$ is allowed at third order in $\lambda$. This means that although the MSG allows magnetically induced electrical polarization along both the [120] and [001] directions, this framework permits to predict a more significant polarization along the [120] direction, without the need to explore specific mechanisms for these magnetoelectric effects.

Coming back to first-order effects, the variation of $\boldsymbol{P}$ induced by collective spin rotation about the [001] axis results to be [57]:

$$P_1^{(1)} = -\lambda\omega_{1,12}^{P,(1)} sin(\varphi), \quad P_2^{(1)} = \lambda\omega_{1,12}^{P,(1)} cos(\varphi), \tag{14}$$

while that of $\boldsymbol{P}$ induced by collective spin rotation about the [100] axis is given by:

$$P_2^{(1)} = \lambda\omega_{1,12}^{P,(1)} cos(\phi), \quad P_3^{(1)} = \lambda\omega_{3,32}^{P,(1)} sin(\phi), \tag{15}$$

where $\varphi$ and $\phi$ are the overall spin rotation angles in the (001) and (100) planes, respectively. It can be seen that, when the spins rotate about the [001] axis, the magnitude of $\boldsymbol{P}$ arising from the lowest-order SOC contribution remains unchanged [Fig. 3(e)]. However, because $\boldsymbol{P}$ in $VCl_2$ is a first-order SOC effect, its vector direction rotates within the plane with the same period as the spin rotation. By contrast, when the spins rotate about the [100] axis, the magnitude of $\boldsymbol{P}$ arising from the lowest-order SOC contribution varies periodically, as shown in Fig. 3(f). This behavior suggests that the in-plane ($P_1$, $P_2$) and out-of-plane electric polarization components have independent origins, corresponding to the different coefficients $\omega_{1,12}^{P,(1)}$ and $\omega_{3,32}^{P,(1)}$ in the SOC expansion. Interestingly, when the spins collectively rotate about the [120] axis, the chirality of the spin spiral configuration remains unchanged, and the lowest-order SOC contribution to the electric polarization maintains the form given in Eq. (13). By contrast, a collective spin rotation of π about either the [100] or [001] axis reverses the spin-spiral chirality and simultaneously reverses the electric polarization. This correspondence directly reflects the locking between the spin spiral configuration and the electric polarization. Moreover, because the electric polarization varies synchronously with the angle of collective spin rotation about [100] and [001] axes, this angular dependence can be used to characterize the orientation of the spin-rotation axis of the spin spiral through polarization measurements.

To further validate the applicability of our approach to real materials, we performed density functional theory (DFT) calculations (See Supplemental Material [57]) for the aforementioned $VSe_2$ and $VCl_2$ systems, obtaining all three components of the electric polarization as a function of spin orientation [Figs. 3(b), 3(c), 3(e) and 3(f)]. The DFT results are in excellent agreement with the SOC expansions derived in Eqs. (11), (12), (14), and (15), confirming that the lowest-order SOC-response terms provide the dominant contributions in real materials. The additional constraints on the electric polarization predicted by our SOC response theory, beyond those imposed by the MSG, are also confirmed by the DFT results.

**c. Relation between the SOC expansion and microscopic magnetoelectric mechanisms**

By treating the SOC tensor as the only symmetry-breaking variable, this method determines the lowest SOC order at which electric polarization is allowed by symmetry, and its dependence on the spin orientation. For example, the OSSG forbids first-order SOC-induced polarization in collinear magnets, while the first- and second-order SOC contributions exhibit different spin-orientation dependences. Meanwhile, the lowest-order SOC contribution to the electric polarization under OSSG constraints may possess fewer degrees of freedom than those permitted by the corresponding MSG, thereby providing additional constraints on the polarization orientation that cannot be obtained from the MSG alone. For example, in $VCl_2$, the ground-state polarization lies predominantly along the [120] direction, with almost negligible MSG-allowed [001] component.

By contrast, microscopic models, such as the inverse Dzyaloshinskii-Moriya (DM) interaction and *p*-*d* hybridization, express the electric polarization in terms of the magnetic configuration $\boldsymbol{M}(\boldsymbol{r}_i)$. At the phenomenological level, these mechanisms correspond to different magnetoelectric coupling terms in a Landau free-energy expansion. The microscopic and SOC expansion approaches therefore classify magnetoelectric effects from distinct perspectives: the former identifies the physical mechanism responsible for the polarization, whereas the latter determines the order of SOC at which this effect emerges. Consequently, there is no universal one-to-one correspondence between an expansion of $\boldsymbol{P}$ in powers of the SOC tensor $\lambda\chi$ and an expansion in terms of $\boldsymbol{M}(\boldsymbol{r}_i)$; rather, their relation depends on the crystal symmetry and atomic arrangement. In this sense, SOC response theory complements conventional microscopic description of magnetoelectric coupling.

Nevertheless, representative connections between the two descriptions can be established for commonly encountered material systems. For collinear systems such as $VSe_2$, the comparison indicates that the contribution associated with *p-d* hybridization first appears at second order in SOC. The Katsura-Nagaosa-Balatsky (KNB) model for the inverse DM mechanism is associated with a Lifshitz-type invariant [15,60] and is generally regarded as a first-order SOC effect. By contrast, SOC-independent mechanisms such as exchange striction correspond directly to a nonzero zeroth-order term $\boldsymbol{\omega}^{P,(0)}$. These relations are summarized in Table I.

The SOC response expansion becomes particularly useful for complex magnetic structures. Whereas identifying microscopic mechanisms directly from the multipolar structure of $\boldsymbol{M}(\boldsymbol{r}_i)$ can rapidly become impractical, the SOC order of each symmetry-allowed contribution follows directly from the OSSG.

## IV. COMPROMISE BETWEEN ANOMALOUS HALL EFFECT AND MAGNETIZATION

Achieving a strong AHE in compensating magnets is one of the goals of next-generation anti-interference magnetic spintronic devices. Numerous attempts have been undertaken in systems such as non-collinear spin chirality AFM [9,10] and altermagnets [39]. However, even when achieving the AHE in these AFMs, net magnetization inevitably persists. This can be easily understood in terms of the MSGs of the structures. Since AHE and magnetization are tensors with the same transformation properties and therefore are subject to the same MSG constraints, symmetry either permits both or forbids both simultaneously. Therefore, a central theme in the study of AHE in AFMs is how to realize a large AHE response in systems with minimal net magnetization.

Recently, the OSSG-based magnetic classification theory-introduced the term spin-orbit magnetism (SOM) to designate a subset of AFM systems exhibiting SOC-induced net magnetization [38]. This classification provides a framework for finding AFM systems with a large AHE response and minimal net magnetization. Specifically, although the AHE and the magnetization transform identically under an MSG, under an OSSG, the AHE shares the same symmetry transformation properties as the orbital magnetization, but they are different from those of the spin magnetization.

Due to their different OSSG constraints, the AHE (together with the orbital magnetization) and the spin magnetization may emerge at different orders in the SOC expansion. Therefore, a material can be designed such that the AHE is a SOC effect of lower order than the spin magnetization. An example of this situation is the case of $Mn_3Sn$ [38]. Given that spin magnetization in most materials significantly outweighs orbital magnetization, attenuating spin magnetization would substantially reduce the net magnetization of the material. This provides a practical route for achieving a large AHE with minimal net magnetization.

**a. Formulations of anomalous Hall effect and magnetization**

Specifically, the intrinsic anomalous Hall conductivity can be expressed as:

$$\sigma_i^{AHE} = \epsilon_{ijk}\frac{e^2}{\hbar}\sum_{n'\neq n}\int_{BZ}\frac{d^3k}{(2\pi)^3}f_{\boldsymbol{k}n}\frac{2Im\left[\left\langle \boldsymbol{k}n\middle|\partial_{k_j}\widehat{H}(\boldsymbol{k})\middle|\boldsymbol{k}n'\right\rangle\left\langle \boldsymbol{k}n'\middle|\partial_{k_k}\widehat{H}(\boldsymbol{k})\middle|\boldsymbol{k}n\right\rangle\right]}{\left(\epsilon_{\boldsymbol{k}n}-\epsilon_{\boldsymbol{k}n'}\right)^2}, \tag{16}$$

where $\epsilon_{ijk}$ is the Levi-Civita symbol; $f_{\boldsymbol{k}n}$ is the Fermi-Dirac distribution, $\widehat{H}(\boldsymbol{k})$ is the system Hamiltonian, and $|\boldsymbol{k}n\rangle$, $|\boldsymbol{k}n'\rangle$ are the eigenstates of the system. From symmetry considerations, the anomalous Hall conductivity vector $\boldsymbol{\sigma}^{AHE} = (\sigma_1^{AHE}, \sigma_2^{AHE}, \sigma_3^{AHE})$ transforms as an axial vector in real space and is odd under time-reversal. Therefore, it shares the same symmetry transformation properties as the orbital magnetization $\boldsymbol{M}_o$, and consequently follows the same expansion form in terms of the SOC tensor. The transformations of $\boldsymbol{\sigma}^{AHE}$, $\boldsymbol{M}_o$ and $\boldsymbol{M}_s$ under the point-group part of an OSSG operation can be expressed as:

$$\boldsymbol{\sigma}^{AHE} \xrightarrow{\{g_s||g_l\}} det(R_s)det(R_l)R_l\boldsymbol{\sigma}^{AHE}, \tag{17}$$

$$\boldsymbol{M}_o \xrightarrow{\{g_s||g_l\}} det(R_s)det(R_l)R_l\boldsymbol{M}_o, \tag{18}$$

$$\boldsymbol{M}_s \xrightarrow{\{g_s||g_l\}} R_s\boldsymbol{M}_s. \tag{19}$$

The Jahn symbols describing the transformation properties under the OSSG of AHE, $\boldsymbol{M}_o$ and $\boldsymbol{M}_s$ are $a\{V^2\}$, $aeV$ and $M$, respectively. Due to the fact that $\{V^2\}$ transforms as an axial vector (i.e., $eV$), $\boldsymbol{\sigma}^{AHE}$ and $\boldsymbol{M}_o$ have the same transformation properties. $\boldsymbol{\sigma}^{AHE}$, $\boldsymbol{M}_o$ and $\boldsymbol{M}_s$ can then be matched with the SOC tensor $\lambda\boldsymbol{\chi}$ to determine their orders of SOC. Similarly to the approach taken for electrical polarization $\boldsymbol{P}$ in Sec. III, we consider the response relations expressed as a series expansion:

$$\sigma_a^{AHE}[\boldsymbol{\chi}] = \omega_a^{o,(0)} + \lambda \sum_{ij} \omega_{a,ij}^{o,(1)} \chi_{ij} + \lambda^2 \sum_{ijkl} \omega_{a,ij,kl}^{o,(2)} \chi_{ij}\chi_{kl} + \cdots, \tag{20}$$

$$M_{s,a}[\boldsymbol{\chi}] = \omega_a^{s,(0)} + \lambda \sum_{ij} \omega_{a,ij}^{s,(1)} \chi_{ij} + \lambda^2 \sum_{ijkl} \omega_{a,ij,kl}^{s,(2)} \chi_{ij}\chi_{kl} + \cdots, \tag{21}$$

where $\boldsymbol{\omega}^{(n)}$ denotes an undetermined coefficient tensor of rank $(2n+1)$; $a$ indicates the components of the vectors $\boldsymbol{\sigma}^{AHE}$, and $\boldsymbol{M}_s$; the indices $i$, $j$, $k$, $l$ label the components of the SOC tensor. Since $\boldsymbol{M}_o$ and $\boldsymbol{\sigma}^{AHE}$ share identical constraints under OSSG, their response tensor forms to the SOC tensor $\lambda\boldsymbol{\chi}$ are also identical, we thus omit the presentation of the formulation of $\boldsymbol{M}_o$. Based on the invariance principle, the relationship between $\boldsymbol{\sigma}^{AHE}$, $\boldsymbol{M}_s$ and $\lambda\boldsymbol{\chi}$ under the constraints of a OSSG point-group operation $\{g_s||g_l\}$ can be derived as:

$$det(R_s)det(R_l)R_l\, \boldsymbol{\sigma}^{AHE}[\boldsymbol{\chi}] = \boldsymbol{\sigma}^{AHE}[det(R_s)det(R_l)R_l\boldsymbol{\chi}R_s^{-1}], \tag{22}$$

$$R_s\boldsymbol{M}_s[\boldsymbol{\chi}] = \boldsymbol{M}_s[det(R_s)det(R_l)R_l\boldsymbol{\chi}R_s^{-1}]. \tag{23}$$

Therefore, the allowed form of the tensor $\boldsymbol{\omega}^{o,(n)}$ of $\boldsymbol{\sigma}^{AHE}/\boldsymbol{M}_o$ in Eq. (20) can be obtained under the symmetry constraints of the OSSG in Eq. (22), which transforms according to the Jahn symbol $aeV \cdot (aeVM)^n$. Similarly, the OSSG constraints on the tensor $\boldsymbol{\omega}^{s,(n)}$ of $\boldsymbol{M}_s$ can be obtained by combining Eqs. (21) and (23), thus the corresponding Jahn symbol is $M \cdot (aeVM)^n$. For example, in first order, the Jahn symbol $V^2M$ for the tensor $\boldsymbol{\omega}^{o,(1)}$ of $\boldsymbol{\sigma}^{AHE}/\boldsymbol{M}_o$ leads to the transformation law:

$$\omega_{nmi}^{o,(1)} = \sum_{kjl}[R_l]_{nk}[R_l]_{mj}[R_s]_{il}\omega_{kjl}^{o,(1)}. \tag{24}$$

whereas the symbol *aeMVM* for the tensor $\boldsymbol{\omega}^{s,(1)}$ of $\boldsymbol{M}_s$ results in

$$\omega_{nmi}^{s,(1)} = \sum_{kjl} det(R_s)det(R_l)[R_s]_{nk}[R_l]_{mj}[R_s]_{il}\omega_{kjl}^{s,(1)}. \tag{25}$$

The invariance of the $\boldsymbol{\omega}^{(n)}$ tensors under these transformations for all the operations of the spin point group corresponding to the OSSG determines the symmetry dictated form of these tensors.

Because the $\boldsymbol{\omega}^{(n)}$ tensors of $\boldsymbol{\sigma}^{AHE}/\boldsymbol{M}_o$ and $\boldsymbol{M}_s$ are subject to different OSSG transformations, the lowest nonzero order $n$ in the expansion of $\boldsymbol{\sigma}^{AHE}/\boldsymbol{M}_o$ and $\boldsymbol{M}_s$ may differ. This distinction provides a symmetry-based guideline for independently tuning the AHE and the net magnetization in real materials.

**b. High-throughput material analysis and interpretation**

After establishing the response relations between the SOC tensor $\lambda\boldsymbol{\chi}$ and $\boldsymbol{\sigma}^{AHE}$, $\boldsymbol{M}_o$ and $\boldsymbol{M}_s$, we can carry out a systematic analysis of real materials retrieved from the MAGNDATA database [61]. This analysis has already been implemented through our online program FINDSPINGROUP [62]. For all identified SOM materials in Ref. [38], we summarize the $\lambda\boldsymbol{\chi}$ expansions of $\boldsymbol{\sigma}^{AHE}/\boldsymbol{M}_o$ and $\boldsymbol{M}_s$ in Supplemental Material Table I [57].

According to the lowest orders in the SOC expansions of $\boldsymbol{\sigma}^{AHE}/\boldsymbol{M}_o$ and $\boldsymbol{M}_s$ (denoted as $i_o$ and $i_s$ in the following), all 224 SOM materials can be classified into three main categories: $i_o < i_s$, $i_o = i_s$, and $i_o > i_s$, as shown in Fig. 4(a). The first category represents the ideal scenario for realizing a large AHE in small magnetization systems, and includes 32 materials. Since higher-order SOC responses generally correspond to smaller magnitudes, this case is the most favorable for balancing the AHE with the magnetizations $\boldsymbol{M}_o$ and $\boldsymbol{M}_s$. In the second category, there are two subcategories, depending on whether both $i_o$ and $i_s$ are of either first or third order. The database contains 182 and 9 materials in these two subcategories, respectively. Notably, cases where both $i_o$ and $i_s$ are first order account for the vast majority in the database. There are no materials for which both $i_o$ and $i_s$ are of the second order at present. Finally, the case in which $i_o > i_s$ currently contains only one material in the database, namely $Ba_3MnSb_2O_9$, whose OSSG is $C^{2_{010}}2/^{2_{001}}c|(3^2_{001},1,1;3^1_{001})^{m_{001}}1$. In this material, $i_o$ is of third order, whereas $i_s$ is of second order.

Next, we further analyzed by induction the 32 materials for which $i_o < i_s$, and found that they can be straightforwardly classified into four types, as shown in Table II. Type I includes 7 materials, all of which exhibit the magnetic geometry characteristics of collinearity (spin-only group ${}^{\infty m}1$) and symmetry {-1||$4^1$}. Under these constraints, $i_o$ is of first order, whereas $i_s$, is second order. The representative material $NiF_2$ is shown in Fig. 4(b). Type II includes 10 materials, all of which possess coplanar magnetic configurations related by the symmetry operation {$3^2$||-$3^1$}. Among them, 9 belong to the Kagome $Mn_3X$ family and its derivatives, where X is a main-group element (e.g., As, Ge, Sn) or a late transition metal (e.g., Ir, Pt). Same as Type I, $i_o$ is first order, whereas $i_s$ is second order. The representative material $Mn_3Sn$ is shown in Fig. 4(c). Type III includes 5 materials, all sharing the same chemical formula $ABC_3$, and OSSG $P^{2_{100}}6_3/^{2_{120}}m^{2_{120}}c^{2_{001}}m^{m_{120}}1$. In this type, $i_o$ is second order, whereas $i_s$ is third order. The representative material $CsCoBr_3$ is shown in Fig. 4(d). Type IV includes 10 noncoplanar materials, corresponding to the case in which a nonzero $\boldsymbol{\sigma}^{AHE}/\boldsymbol{M}_o$ exists even without SOC. Accordingly, $i_o$ is zeroth order, whereas $i_s$ is of first order. This case is allowed only when a noncoplanar magnetic structure breaks the spin-only groups ${}^{m}1$ or ${}^{\infty m}1$. The representative material $CoNb_3S_6$ is shown in Fig. 4(e).

Overall, each type exhibits highly specific characteristics, indicating that the condition $i_o < i_s$ is not a common feature, but instead requires particular symmetry conditions. A systematic study of the symmetry breaking induced by SOC provides precisely this distinctive perspective. In addition, it is worth noting that, among all these materials, the difference between $i_o$ and $i_s$ never exceeds one order. This relation can also be understood directly from symmetry considerations. The Jahn symbol of the *i*-th order coefficient $\boldsymbol{\omega}^{o,(i)}$ for $\boldsymbol{\sigma}^{AHE}/\boldsymbol{M}_o$ is $eaV(eaVM)^i$, whereas that of the *i*-th order coefficient $\boldsymbol{\omega}^{s,(i)}$ for $\boldsymbol{M}_s$ is $M(eaVM)^i$. By exchanging the index ordering of the magnetization and the $\lambda\boldsymbol{\chi}$ series in *i*-th order orbital coefficient $\boldsymbol{\omega}^{o,(i)}$ and the (*i+1*)-th order spin coefficient $\boldsymbol{\omega}^{s,(i+1)}$, respectively, their Jahn symbols can be rewritten as $(eaVM)^i eaV$ and $(eaVM)^i eaVMM$, which differ only by the factor $MM = M^2$.

Because the permutation of tensor indices does not affect whether a given tensor vanishes, and $M^2$ always contains the identity representation of any group, an allowed $i$-th order contribution to $\boldsymbol{\sigma}^{AHE}/\boldsymbol{M}_O$ necessarily implies an allowed *(i+1)*-th order contribution to $\boldsymbol{M}_S$.

In addition to the above analysis comparing the orders of $\boldsymbol{\sigma}^{AHE}/\boldsymbol{M}_O$ and $\boldsymbol{M}_S$, we also note that in 10 materials, the directions of $\boldsymbol{\sigma}^{AHE}$/net magnetization $\boldsymbol{M}$ allowed by the MSG differ from those permitted by the lowest-order SOC terms under OSSG constraints. Since the forms of physical quantities allowed by the MSG represent the combinations of all orders in the SOC expansion, this discrepancy implies that different components can originate at different SOC orders, as demonstrated in the previous section for the case of electrical polarization. Therefore, although the possible directions of $\boldsymbol{\sigma}^{AHE}/\boldsymbol{M}_O$ in real materials are dictated by the MSG, a more restrictive constraint resulting from the lowest-order SOC term may in practice exist. This is precisely the additional information provided by the SOC response expansion under OSSG symmetry.

## V. DISCUSSION

The SOC response theory incorporating OSSGs presented here in detail makes it possible to separate the symmetry breaking associated with magnetic order from that introduced by SOC. The magnetic state in the absence of SOC is taken as the reference state, while the SOC tensor is treated as the only additional symmetry-breaking variable. This formulation enables the systematic identification of physical responses that originate from SOC, independently of the symmetry reduction already imposed by magnetic order.

Because the SOC-free magnetic symmetry is fully characterized by the OSSG, the symmetry analysis depends only on the SOC tensor and does not require explicit treatment of the Wyckoff positions of the magnetic atoms. More generally, our results show that the OSSG is useful not only for characterizing SOC-free properties, as has been primarily emphasized previously, but also for analyzing SOC-induced phenomena and their dependence on the spin orientation. This approach therefore completes the symmetry-based description of magnetic systems.

In summary, the present work establishes a general symmetry framework for the analysis of SOC-induced response tensors in magnetic materials, by combining the symmetry description of magnetic order provided by the OSSG with the symmetry breaking introduced by the SOC tensor. Although demonstrated here for the electric polarization, the anomalous Hall effect, and the spin magnetization, the same procedure can be applied to any SOC-driven physical tensor property. Through the generalized Jahn symbol description, the symmetry-adapted forms of response tensors for any physical property can be systematically obtained at each order in the SOC expansion using the program STENSOR, recently made available at the Bilbao Crystallographic Server [63]. This framework is therefore expected to facilitate future investigations into a broad range of SOC-related phenomena, including the spin Hall effect, the spin-orbit torque, and the nonlinear Hall effect.

**Acknowledgements**

This work was supported by National Natural Science Foundation of China (12525410, 12274194, 12574275 and 12534003), the National Key R&D Program of China (2025YFA1411300), Guangdong Provincial Quantum Science Strategic Initiative (GDZX2601003 and GDZX2401002), Shenzhen Science and Technology Program (RCJC20221008092722009, 20231117091158001 and QNXMC20250701094702004), the Innovative Team of General Higher Educational Institutes in Guangdong Province (2020KCXTD001) and Center for Computational Science and Engineering of Southern University of Science and Technology.

Table I. SOC expansion and representative microscopic magnetoelectric mechanisms in Type-II multiferroics. The electric polarization is classified according to the lowest order in SOC at which it appears, together with the corresponding representative microscopic mechanisms and material examples.

| Lowest nonzero order | SOC expansion | Representative mechanism | Material examples |
|---|---|---|---|
| 0 | $\omega_a^{P,(0)} + \cdots$ | Exchange striction | $HoMnO_3$, $GdMn_2O_5$ |
| 1 | $\lambda \sum_{ij} \omega_{a,ij}^{P,(1)} \chi_{ij} + \cdots$ | Inverse DM-type mechanism | Spiral $VCl_2$, $VBr_2$ |
| 2 | $\lambda^2 \sum_{ijkl} \omega_{a,ij,kl}^{P,(2)} \chi_{ij} \chi_{kl} + \cdots$ | *p*-*d* hybridization | Collinear $VSe_2$ |
| … … | | | |

Table II. The four types of materials identified having the AHE/orbital magnetization ($\boldsymbol{\sigma}^{AHE}$ / $\boldsymbol{M}_o$) as a SOC lower-order effect (or SOC-free) compared to the spin magnetization ($\boldsymbol{M}_s$). The corresponding magnetic geometric features, the lowest allowed orders of SOC for both orbital and spin contributions ($i_o$ and $i_s$, respectively), and material examples are listed.

| Types | Magnetic geometry character | $i_o$ | $i_s$ | Material examples |
|---|---|---|---|---|
| I | Collinear $\{-1\|\|4^1\}$ | 1 | 2 | $NiF_2$, $SrMn_2V_2O_8$ |
| II | Coplanar $\{3^2\|\|-3^1\}$ | 1 | 2 | $Mn_3NiN$, $Mn_3Sn$ |
| III | $P^{2_{210}}6_3/^{2_{010}}m^{2_{010}}c^{2_{001}}m^{m_{010}}1$ | 2 | 3 | $CsCoBr_3$, $CsNiCl_3$ |
| IV | Noncoplanar | 0 | 1 | $DyCrWO_6$, $CoNb_3S_6$ |

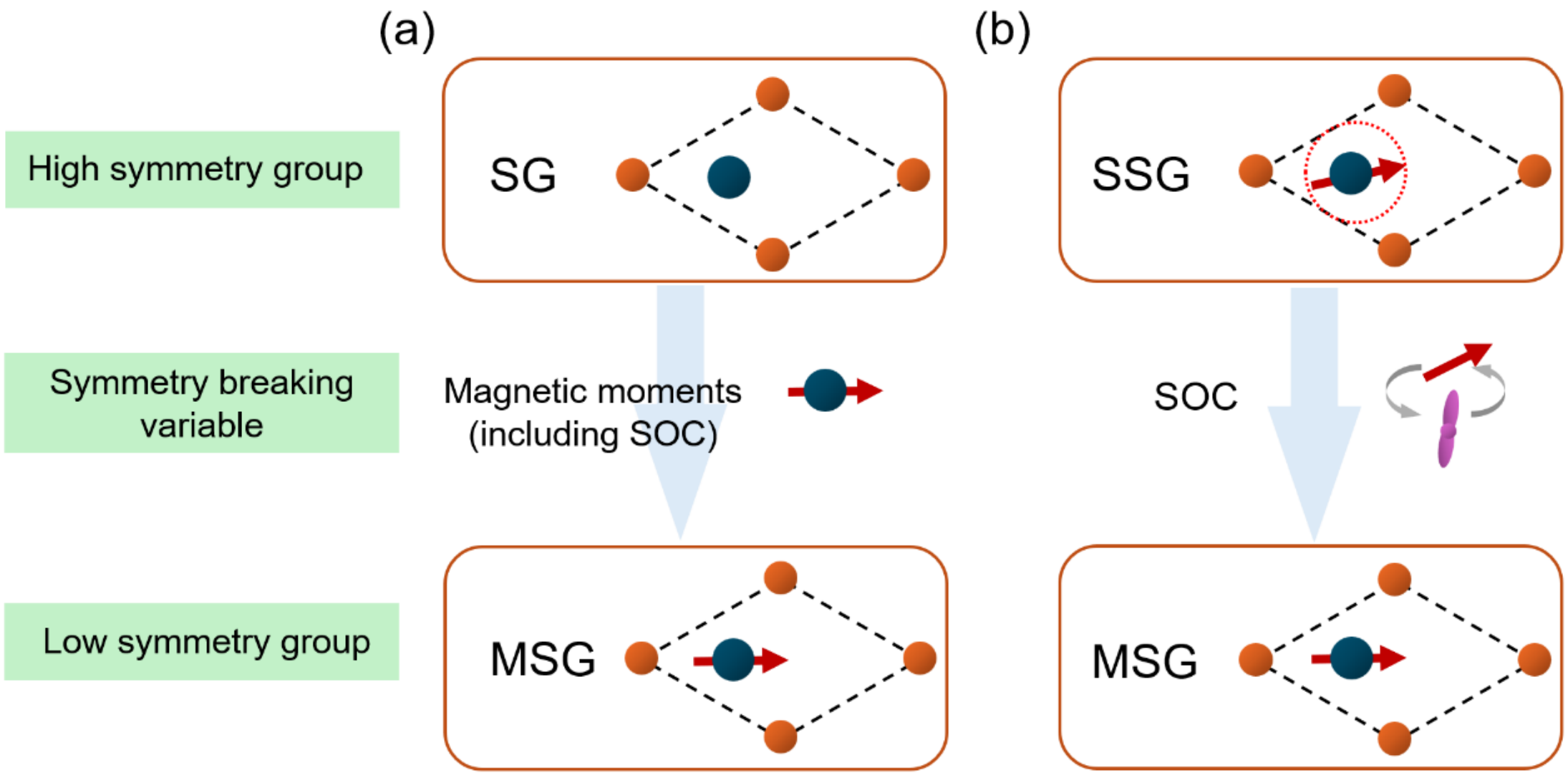

Fig. 1. Comparison between (a) the traditional symmetry-breaking theory including spin-orbit coupling (SOC), and (b) the present approach, which treats SOC as an additional symmetry-breaking perturbation. In (a), the magnetic moments, which include SOC, serve as the symmetry breaking variables, leading to a reduction from the space group (SG) to the magnetic space group (MSG). In (b), SOC itself serves as an additional symmetry-breaking perturbation, leading to a reduction from the spin space group (SSG) of the primary magnetic order to the MSG. Blue and orange spheres represent magnetic and non-magnetic atoms, respectively, and red arrows denote spins.

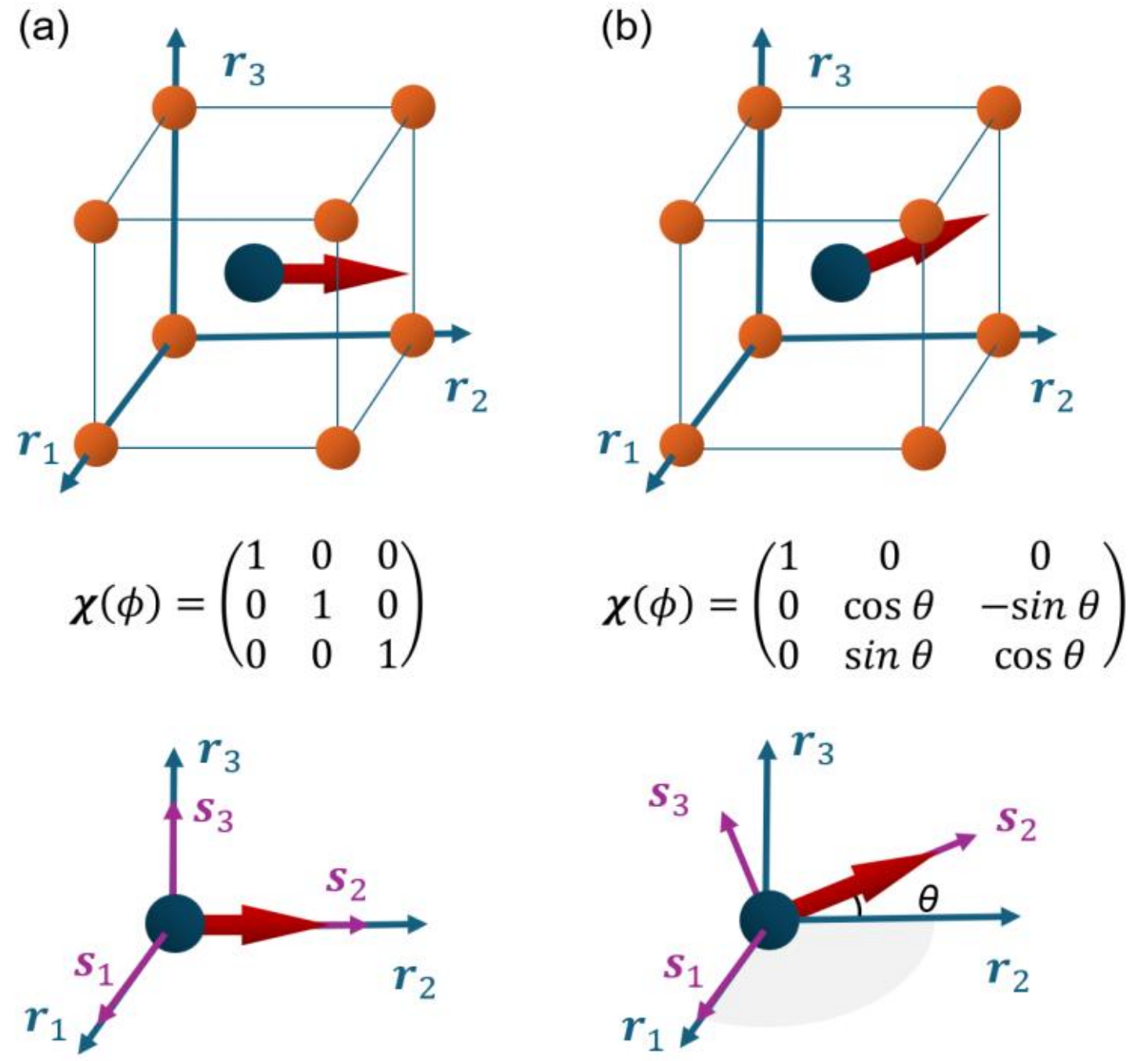

Fig. 2. The form of the SOC tensor in the magnetic ground state and during a collective spin rotation with respect to the magnetic ground state. (a) The magnetic ground state with the real-space and spin-space coordinate bases, $\boldsymbol{r}_i$ and $\boldsymbol{s}_j$, respectively, following the OSSG standardization. (b) The state obtained by a collective spin rotation about axis $\boldsymbol{r}_1$ by an angle $\theta$ with respect to the magnetic ground state. The blue and orange spheres represent magnetic and non-magnetic atoms, respectively. The red arrows represent the spins.

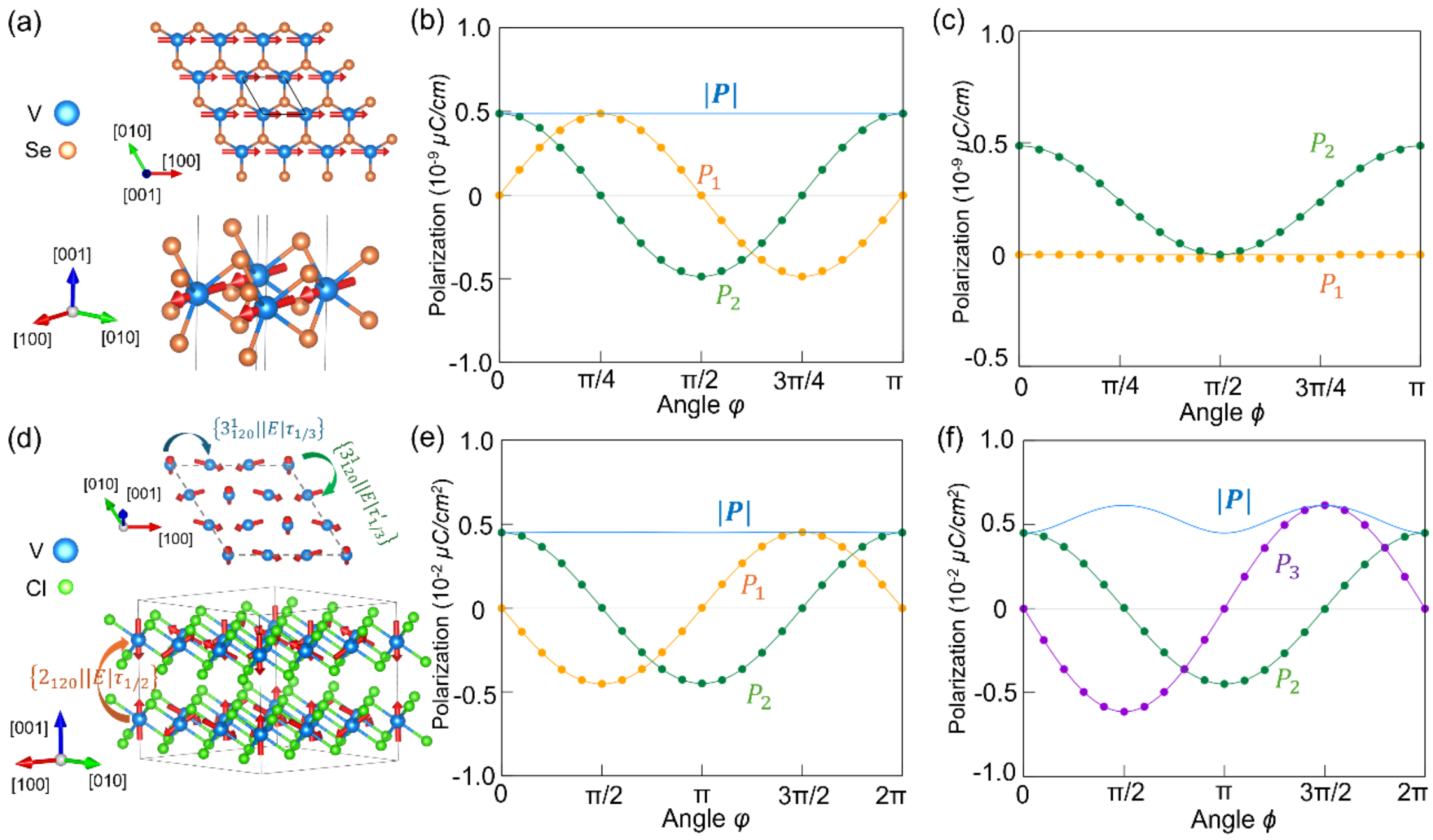

Fig. 3. The dependence of the electric polarization on the spin orientation in $VSe_2$ and $VCl_2$. (a) Magnetic structure of $VSe_2$. (b-c) Variation of the magnitude of the electrical polarization ($|\boldsymbol{P}|$) and its three components ($P_1$, $P_2$, $P_3$) in (b) the (001) plane and (c) the (120) plane. (d) Magnetic structure of $VCl_2$. The spin translational operations characterizing the spin spiral configuration are indicated in the figure. (e-f) Variation of the magnitude of electrical polarization ($|\boldsymbol{P}|$) and its three components ($P_1$, $P_2$, $P_3$) in (e) the (001) plane and (f) the (100) plane. The scale on the vertical axis of the electric polarization plot represents the ratio of the magnitude of the electric polarization to that of the ground state ($\varphi = 0$, $\phi = 0$). The points represent the results of DFT calculations, while the lines represent the results obtained from the SOC response theory presented in this work.

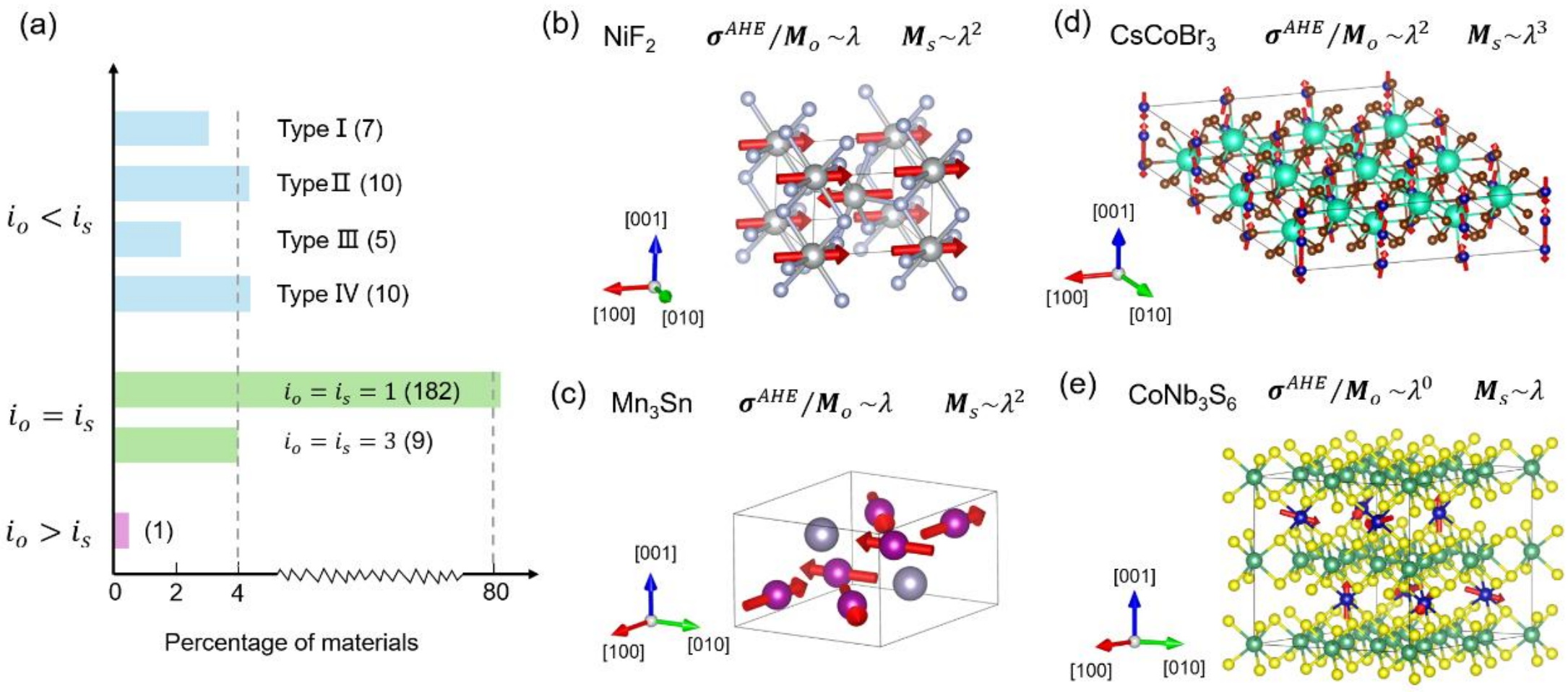

Fig. 4. Classification of SOM materials according to the lowest OSSG-allowed SOC orders of $\boldsymbol{\sigma}^{AHE}/\boldsymbol{M}_o$ and $\boldsymbol{M}_s$. (a) Statistics of the seven cases of SOM materials according to the relation between $i_o$ and $i_s$ from the MAGNDATA database. $i_o$ and $i_s$ denote the lowest OSSG-allowed SOC orders of $\boldsymbol{\sigma}^{AHE}/\boldsymbol{M}_o$ and $\boldsymbol{M}_s$, respectively. (b-e) Magnetic structure and lowest OSSG-allowed SOC orders of $\boldsymbol{\sigma}^{AHE}/\boldsymbol{M}_o$ and $\boldsymbol{M}_s$ of (b) $NiF_2$, (c) $Mn_3Sn$, (d) $CsCoBr_3$, (e) $CoNb_3S_6$.